\documentclass{svmult}

\usepackage{graphicx}        
\usepackage[dcucite]{harvard} 

\usepackage{color}

\begin{document}

\title*{Blackouts, risk, and fat-tailed distributions}

\author{Rafa{\l} Weron\inst{1} and  Ingve Simonsen\inst{2}}

\institute{Hugo Steinhaus Center for Stochastic Methods,\\
    Wroc{\l}aw University of Technology, 50-370 Wroc{\l}aw, Poland
\and 
    Department of Physics, NTNU, NO-7491 Trondheim, Norway} 

\maketitle

\setcounter{footnote}{0}


\begin{abstract}
  We analyze a 19-year time series of North American electric power
  transmission system blackouts.  Contrary to previously reported
  results we find a fatter than exponential decay in the distribution
  of inter-occurrence times and evidence of seasonal dependence in the
  number of events. Our findings question the use of self-organized
  criticality, and in particular the sandpile model, as a paradigm 
  of blackout dynamics in power transmission systems. Hopefully, though, 
  they will provide guidelines to more accurate models for evaluation 
  of blackout risk.  
\end{abstract}



Electric power transmission networks are complex systems.\footnote{For
  a brief review of approaches to complex systems and cascading
  failure in power system blackouts see
  \citeasnoun{dob:car:lyn:new:04}.} Due to economic factors, they are
commonly run near their operational limits. Major cascading
disturbances or blackouts of these transmission systems have serious
consequences. Although, each blackout can be attributed to a
particular cause: natural peril, equipment malfunction or human
behavior, an exclusive focus on the causes of these events can
overlook the global dynamics of a complex system. Instead, it might be
interesting to study blackouts from a top-down perspective. Following
\citeasnoun{car:new:dob:poo:04} we analyze a time series of blackouts
to explore the nature of these complex systems. However, despite the
fact that we are using the same database we obtain different results.
Consequently, we challenge their arguments that lead to modeling
blackouts as a self-organized criticality (SOC) phenomenon
\cite{bak:tan:wie:87}.



The reliability events --- like the August 1996 blackout in
Northwestern America that disconnected 30,390 MW of power to 7.5
million customers or the even more spectacular August 2003 blackout in
Northeastern America that disconnected 61,800 MW of power to 50
million people --- demonstrate that the necessary operating practices,
regulatory policies, and technological tools for dealing with the
changes are not yet in place to assure an acceptable level of
reliability. In a restructured environment, prices are a matter of
private choice, yet the reliability of the delivery system affects everyone. 

Naturally, the operation of the electric system is more difficult to
coordinate in a competitive environment, where a much larger number of
parties are participating. For example, in North America about
one-half of all domestic generation is now sold over ever-increasing
distances on the wholesale market before it is delivered to customers
\cite{alb:alb:nak:04}. Consequently the power grid is witnessing power
flows in unprecedented magnitudes and directions. 
Unfortunately, it seems that the development of reliability management
reforms and operating procedures has lagged behind economic reforms in
the power industry. In addition, responsibility for reliability
management has been disaggregated to multiple institutions
\cite{car:etal:00}. All this results in an increase of the risk of
blackouts, not only in North America, but also world-wide. 


The Disturbance Analysis Working Group (DAWG)
database\footnote{Publicly available from
  http://www.nerc.com/$\tilde{~}$dawg/database.html.} summarizes
disturbances that have occurred in the electric systems in North America. The
database is based on major electric utility system disturbances
reported to the U.S. Department of Energy (DOE) and the North American
Electrical Reliability Council (NERC). The data arise from government
incident reporting requirements criteria detailed in DOE form EIA-417. 

\citeasnoun{car:new:dob:poo:04} analyzed the first 15 years of data
(1984-1998) from the DAWG database. As currently four more years of
data are available\footnote{The delay in data distribution is due to
  the complexity of the problem. It can take months after a large
  blackout to dig through the records, establish the events occurring
  and reproduce a causal sequence of events.} we study two datasets:
D98 covering the period 1984-1998 and D02 covering
the full data set 1984-2002. The first one is used for comparison with
the previous findings, while the second lets us extend the analysis
and draw more up-to-date conclusions. The data are of diverse
magnitude and of varying causes (including natural perils, human
error, equipment malfunction, and sabotage). It is not clear how
complete these data are, but it seems to be the best-documented source
for blackouts in the North American power transmission system. Besides
the date and the region of occurrence, two measures of the event's
severity are given: the amount of power lost (in MW) and the number of
customers affected.

There are 435 documented blackouts in the first 15 years (dataset
D98), which gives on average 29 blackouts per year. A few events have
missing data in one or both of the severity fields. For the analysis
of blackout sizes we have used only those 427 occurrences which have
complete data in both columns.\footnote{However, for the waiting time
  distribution analysis we have used all occurrences. A preprocessed, 
  spreadsheet-ready ASCII format datafile is available from 
  http://www.im.pwr.wroc.pl/$\tilde{~}$rweron/exchlink.html.} The average
inter-occurrence time is 12.6 days, but the blackouts are distributed
over the 15 years in a non-uniform manner with a maximum waiting time
of 252 days between event origins. Furthermore, the mean and the maximum 
restoration times are 14 hours and 14 days, respectively, indicating that 
the inter-occurrence times are more or less equivalent to the quiet times 
(the lapses of time between the end of a blackout and the beginning of 
the next one).

\begin{figure}[tbp]
\begin{center}
\leavevmode
\includegraphics*[width=.8\columnwidth]{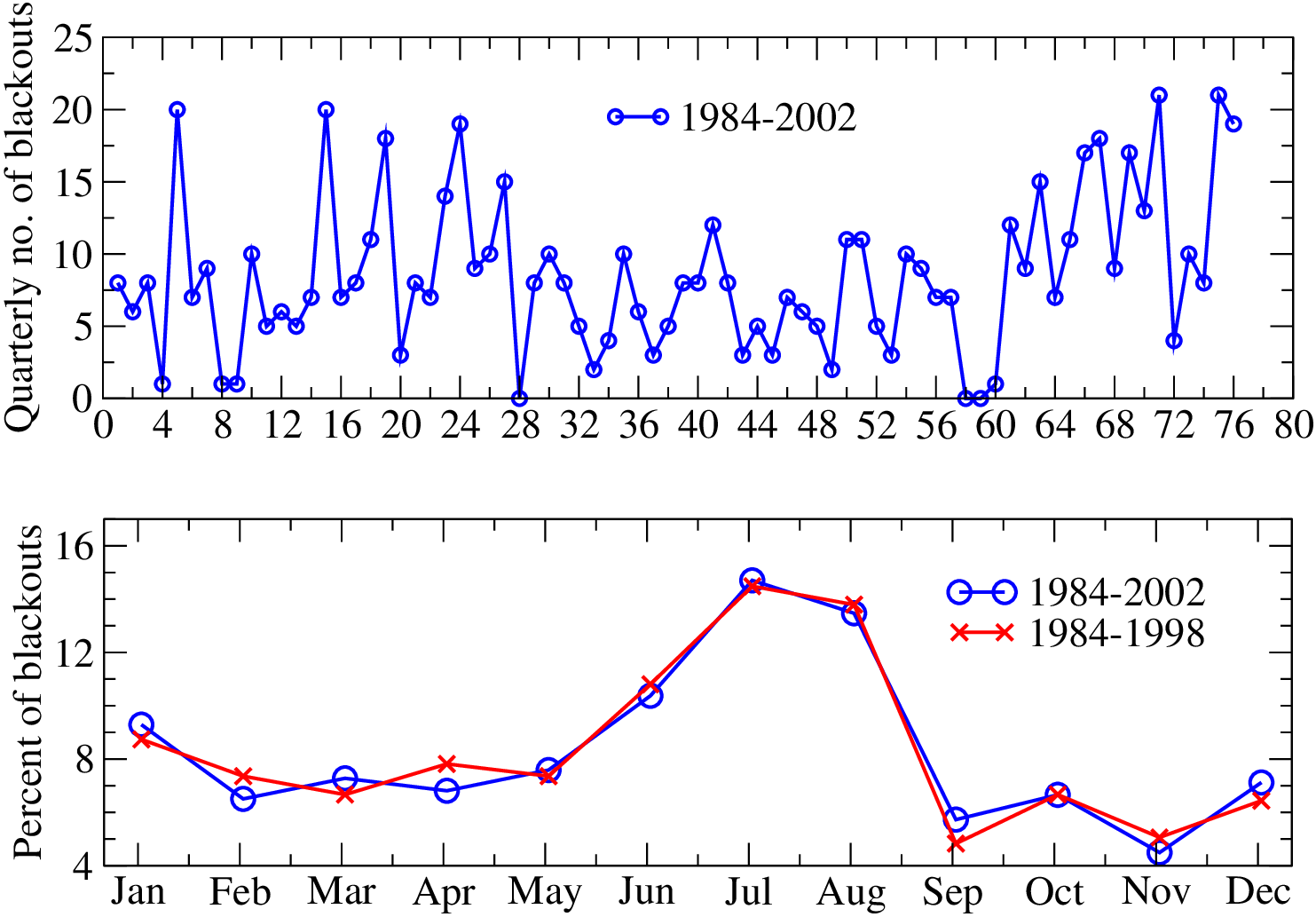}
\caption{The quarterly number of blackout events from 1984 till 2002
  (top) and annual distribution of monthly events (bottom) for the
  North-American power grid.} 
\label{fig-MonthlyEvents}
\end{center}
\end{figure}

In the full dataset (D02) there are 646 documented blackouts, yielding
on average 34 blackouts per year. However, only 578 occurrences have
complete severity data, since -- especially in 1999 and 2000 -- there are many
missing values. The average period of time between blackouts is now
only 10.7 days, indicating a recent increasing trend in the number of
blackouts, while the mean and the maximum restoration times are  
slightly higher: 16 hours and 15 days, respectively.   

Although the scarcity of data limits sound statistical
inference, looking at the top panel of Fig. \ref{fig-MonthlyEvents} we
can intuitively divide the dataset into three parts: an initial period
of relatively volatile activity (1984-1990; quarters 1-28), followed by a fairly calm
period (1991-1998; quarters 29-60), and, most recently, a period of increasing
activity (1999-2002; quarters 61-76). Whether this is a consequence of deregulation,
different incident reporting procedures or simply randomness remains
an open question. However, the seasonal behavior of the outages is
indisputable. Roughly 30\% of all blackouts take place in July and
August, see the bottom panel of Fig. \ref{fig-MonthlyEvents},
regardless of the dataset analyzed.  Our observations contradict
earlier reports, where the authors detected no evidence of systematic
changes in the number of blackouts or (quasi-)periodic behavior
\cite{car:new:dob:poo:04}. 

\begin{figure}[tbp]
\begin{center}
\leavevmode
\includegraphics*[width=.8\columnwidth]{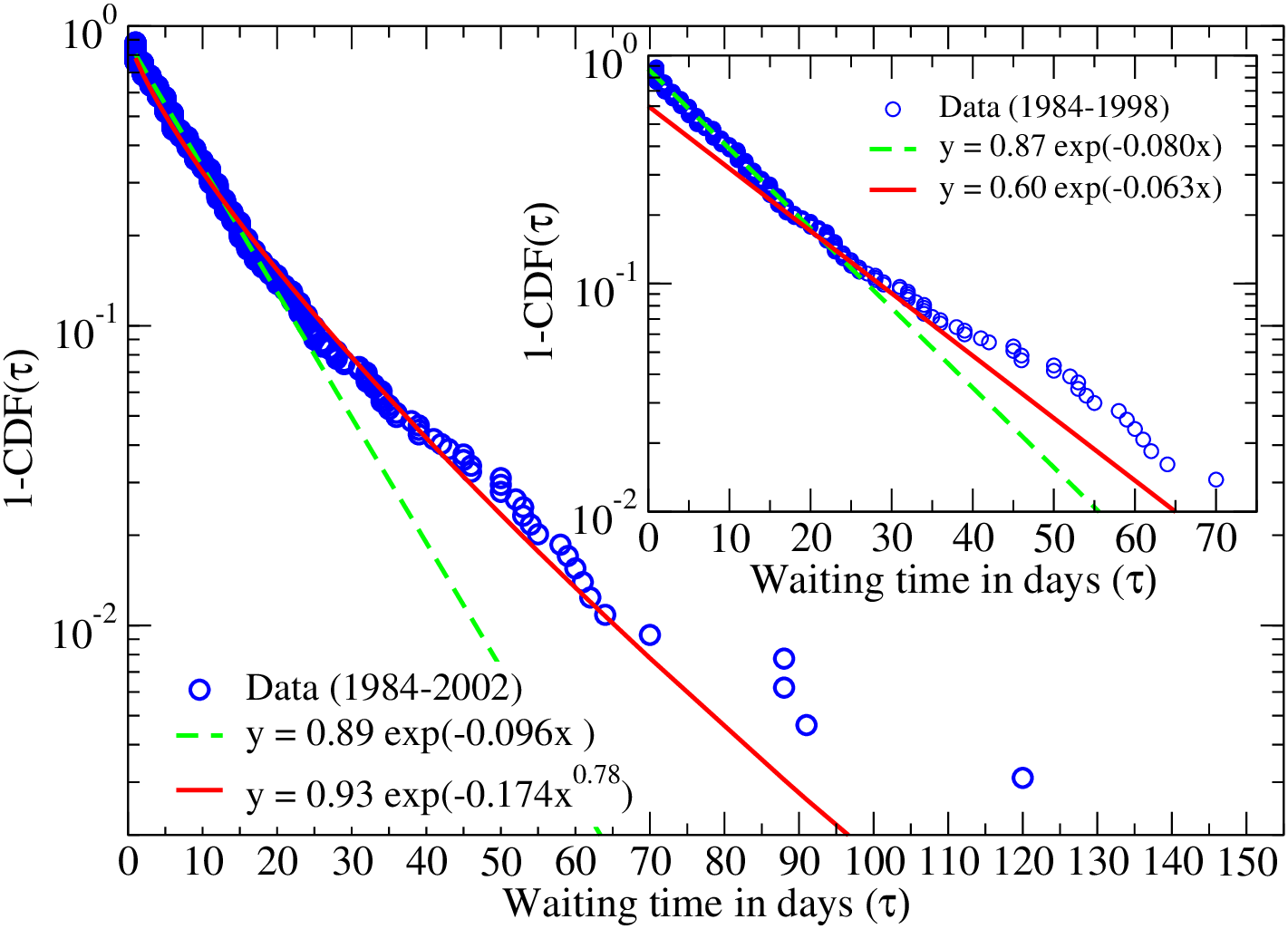}
\caption{The complementary cumulative distribution function
  ($1-\mbox{CDF}(\tau)$) of the waiting times $\tau$ (measured in days)
  between two consecutive blackout origins for the North-American power
  transmission system using the D02 (main panel) and D98 data sets
  (inset).  The dashed lines represent exponential fits to the
  distributions. The solid lines correspond to a stretch exponential
  fit (main panel) and the exponential fit obtained
  by~\protect\citeasnoun{car:new:dob:poo:04} using the same data set
  (inset).  
  } 
\label{fig-WaitingTimes}
\end{center}
\end{figure}

A closer inspection of the waiting times between blackouts reveals a
non-trivial nature. The distribution does not have an exponential
tail, as reported e.g. by \citeasnoun{che:tho:par:01}, but rather a
fatter one.\footnote{Waiting time distribution of high-frequency
  financial data show similar fatter-than exponential distributions
  \protect\cite{scal:gor:mai:man:rab:05}.}  As can be seen in Fig.
\ref{fig-WaitingTimes} the deviation is significant for both D98 and
D02. These findings question the SOC-type approach to modeling
blackout dynamics~\cite{car:new:dob:poo:04} since SOC-type dynamics
should exhibit exponential decay in the waiting time
distribution~\cite{bof:etal:99,car:new:dob:poo:04}.

\begin{figure}[tbp]
\begin{center}
\leavevmode
\includegraphics*[width=.8\columnwidth]{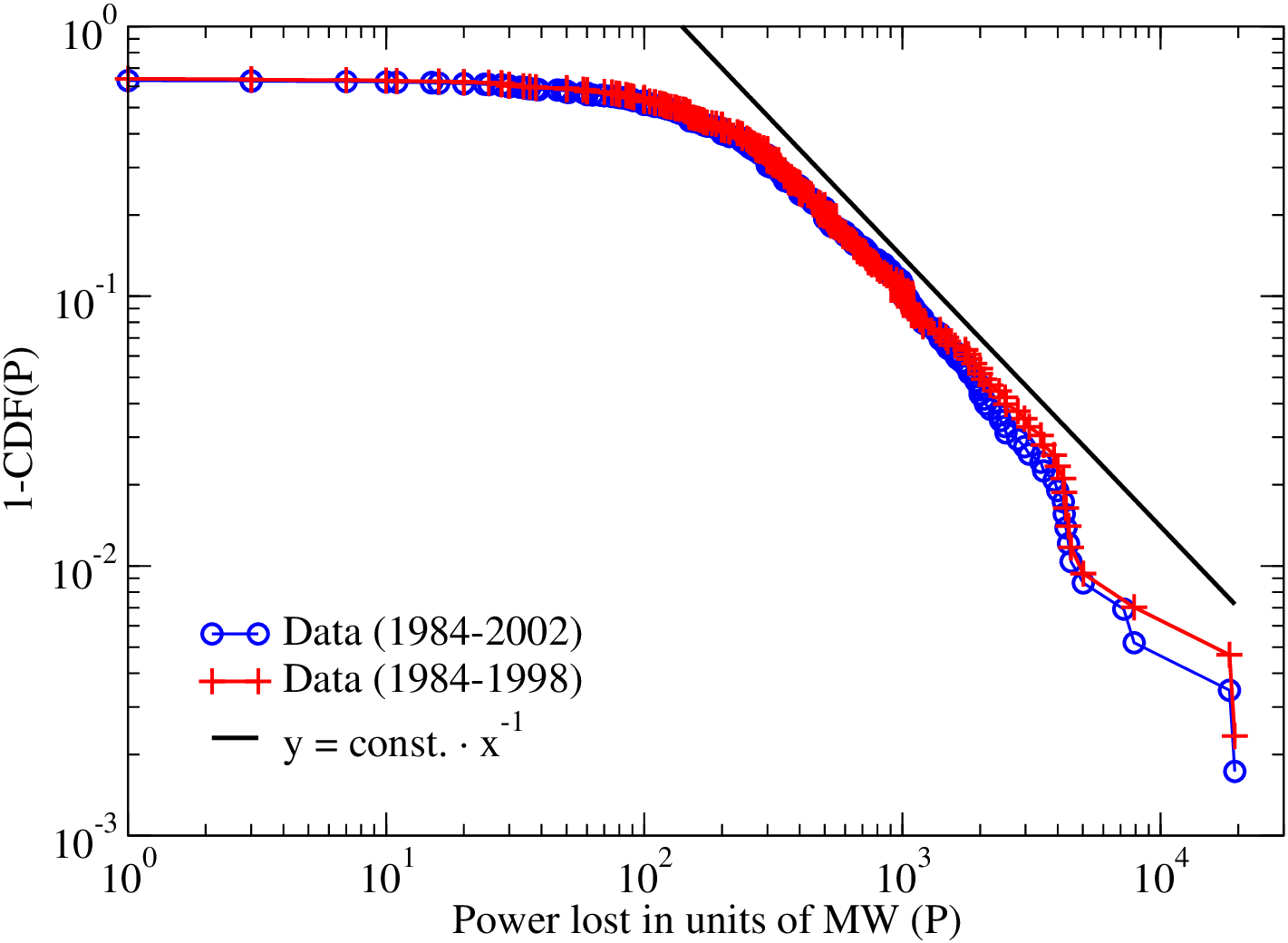}
\caption{The complimentary cumulative distribution ($1-\mbox{CDF}(\mbox{P})$) of
  power lost (P) due to blackouts  for the North-American electric
  power transmission system.} 
\label{fig-PowerLost}
\end{center}
\end{figure}

It is apparent that large blackouts, as the mentioned earlier August
1996 and August 2003 events, are rarer than small blackouts. But how
much rarer are they? Analysis of the D98 and D02 datasets shows that
the complementary cumulative probability distribution of the blackout
sizes does not decrease exponentially with the size of the outage, but
rather has a power-law tail of exponent $\alpha=1$, see 
Fig.\ref{fig-PowerLost}. 
Hence, if we evaluate the risk of a blackout as the product of its frequency 
and cost (commonly regarded to be proportional to unserved energy, see e.g. 
\citeasnoun{bil:all:96}), then the total risk associated with the
large blackouts is -- due to the power-law type distribution of
blackout sizes -- much greater than the risk of small outages. 
This is strong motivation for investigating the global dynamics of series of
blackouts that can lead to power-law tails. The investigated models,
though, should take into account all or at least most of the
characteristics revealed in this study.

\def\bibsection{

\end{document}